\documentstyle[preprint,aps]{revtex}
\draft
\begin{document}
\begin{center}
\noindent {\bf THE REPRESENTATION TRANSFORMATION OF MULTIQUARK WAVE
 FUNCTIONS}
\end{center}
\mbox{}
\begin{center}
Jia-lun Ping$^{a}$, Fan Wang$^{b}$, and T. Goldman$^{c}$
\end{center}

\mbox{}

\noindent $^{a}$ Department of Physics, Nanjing Normal  
University, Nanjing, 210097, China

\noindent $^b$ Center for Theoretical Physics, Nanjing University,  
Nanjing, 210008, China

\noindent $^{c}$ Theoretical Division, Los Alamos National  
Laboratory, Los Alamos, NM 87545, \\
\indent USA
\mbox{}\\
\mbox{}
\begin{center}
{\bf ABSTRACT}\\
\end{center}
It is shown that the representation transformations of multiquark wave
functions between different coupling schemes are just the Racah
coefficients of the permutation group. The transformation coefficients
between the flavor-spin (FS) and the color-spin (CS) schemes are
obtained. As an example, the expansion of the physical bases in terms
of symmetry bases in the CS scheme are given for the interesting cases
$YIJ = 201, 210, 000$.\\
\mbox{}\\
\mbox{}
\newpage

\vspace{0.5cm}

{\noindent {\large {\bf I. Introduction}}}
\vspace{0.5cm}

To understand baryon-baryon interactions, and in turn to search for
dibaryon candidates from the fundamental strong interaction theory QCD,
is still a challenge to contemporary physics. Due to the complexity of
confinement, one relies mainly on QCD inspired models. The central
problem in the computations is the many-body matrix element
calculation. If one wants to do systematic studies, a powerful method
is indispensable.

In the calculation of baryon-baryon interactions and dibaryon masses,
of the bases that can be used to span the model Hilbert space$^{1,2}$,
two are of particular interest. The first is the so-called physical
basis, the other, the symmetry basis. 

The physical basis is nothing more than the quark cluster model basis,
which is constructed directly from two single baryons:
\begin{eqnarray}
\Psi_{\alpha \kappa}(B_1B_2) & = & {\cal A} \left[ \psi(B_1) \psi(B_2) 
\right] ^{[\sigma ]IJ}_{WM_IM_J}  \nonumber \\
 & = & {\cal A} \left[ \left| \begin{array}{c} [\nu_1] \\ 
 ~[\sigma_1] [\mu_1][f_1]Y_1I_1J_1 \end{array} \right\rangle 
  \left| \begin{array}{c} [\nu_2] \\ 
 ~[\sigma_2] [\mu_2][f_2]Y_2I_2J_2 \end{array} \right\rangle \right]
 ^{[\sigma ]IJ}_{WM_IM_J} ,
\end{eqnarray}
where $\alpha=(YIJ)$, $\kappa$ represents all the other quantum numbers
and ${\cal A}$ is the pairwise quark antisymmetrization operator.  The
physical basis has the advantage that it has definite dibaryon content
in the asymptotic region, but it has the disadvantage that it is not
convenient for the calculation of matrix elements. In order to take
advantage of the fractional parentage expansion technique, which has
been widely used in atomic and nuclear physics, one has to use the
symmetry basis. 

The symmetry basis is the group chain classification basis:
\begin{equation}
\Phi_{\alpha K}(q^6) = \left| 
\begin{array}{c}
[\nu]l^3r^3 \\ 
\end{array}
\right\rangle . 
\end{equation}
where $K$ represents other quantum numbers and the group chain we use is
\begin{eqnarray}
\mbox{SU}(36)\supset \mbox{SU}^x(2)\times (\mbox{SU}(18)\supset 
\mbox{SU}^c(3)\times (\mbox{SU}(6) \nonumber \\
\supset ( \mbox{SU}^f(3)\supset \mbox{SU}^{\tau}(2)\times \mbox{U}^Y(1)
) \times \mbox{SU}^{\sigma}))(2)
\end{eqnarray}
\noindent The symmetry basis has its own disadvantage: Most of the
quantum numbers in Eq.(3) are just mathematical labels with no physical
meaning.  Thus they do not describe either conserved quantum numbers,
nor definite dibaryon content in the asymptotic region.

To take both the advantages of these two bases, one must first express
the physical basis in terms of symmetry basis, then calculate the
many-body matrix element using the symmetry basis, and finally
transform the results back to the physical basis. Hence, the
transformation between the physical and symmetry bases is needed. Since
the two bases both are complete in the truncated model Hilbert space,
they are related by a unitary transformation.  The needed
transformation coefficients in the $u$, $d$ and $s$ 3-flavor world have
been tabulated for the flavor-spin (FS) scheme$^{1,3}$ specified by
Eq.(3).

In different applications, the coupling orders are also different. In the FS
scheme, the coupling order is as follows:

(1) $SU^{f\sigma}(6) \supset SU^f(3) \times SU^{\sigma}(2), \hspace{0.5in}
 [f] \times [\sigma_J] \rightarrow [\mu] $.\\

(2) $SU^{cf\sigma}(18) \supset SU^c(3) \times SU^{f\sigma}(6), \hspace{0.5in}
 [\sigma] \times [\mu] \rightarrow [{\tilde \nu}] $.\\

(3) $SU^{xcf\sigma}(36) \supset SU^x(2) \times SU^{cf\sigma}(18),
 \hspace{0.5in} [\nu] \times [{\tilde \nu}] \rightarrow [1^6] $.\\

\noindent In the most popular color-spin (CS) scheme, the coupling order is:

(1) $SU^{c\sigma}(6) \supset SU^c(3) \times SU^{\sigma}(2), \hspace{0.5in}
 [\sigma] \times [\sigma_J] \rightarrow [\mu'] $.\\

(2) $SU^{c\sigma f}(18) \supset SU^{c\sigma}(6) \times SU^f(3), \hspace{0.5in}
 [\mu'] \times [f] \rightarrow [{\tilde \nu}] $.\\

(3) $SU^{xc\sigma f}(36) \supset SU^x(2) \times SU^{c\sigma f}(18), 
 \hspace{0.5in} [\nu] \times [{\tilde \nu}] \rightarrow [1^6] $.\\

Each of these schemes has its advantages. For example, in the CS 
scheme, the expectation values of the color magnetic operator
$$ 
\sum_{i<j} {\bf \lambda_i \cdot \lambda_j} {\bf \sigma_i \cdot \sigma_j} 
$$
can be easily calculated. In the FS scheme useful results already exist.

To understand the relation between results obtained in different
coupling schemes and make our Tables$^3$ more useful (for example, to
expand the physical bases in terms of symmetry bases in the CS scheme),
it is necessary to know the unitary representation ({\it rep})
transformation between the bases in the different schemes.  Fl.
Stancu$^4$ has studied the {\it rep} transformation in the $u$, $d$
2-flavor world and obtained the transformation coefficients for the
cases ($YIJ$) = (201) and (200) as well as several trivial cases.

Because of the plethora of phase conventions possible for the
Clebsch-Gordon (CG) coefficients of SU($n$) for $n\geq 3$, it is highly
desirable to have a systematic and phase consistent method to calculate
the {\it rep} transformation coefficients. In this work, we will prove
that the {\it rep} transformation coefficients are just the Racah
coefficients of permutation group, $S(n)$. (See Sec. II). Since the
S($n$) $\supset$ S($n-1$) isoscalar factors are available$^5$, the
Racah coefficients of S($n$) for $n\leq 6$ can be calculated based on
the genealogical method. As an application, the expansion of the
physical bases in terms of symmetry bases in the CS scheme are
obtained. (See Sec. III).

\vspace{1cm}

{\noindent {\large {\bf II. Rep transformation coefficients and Racah 
coefficients of S($n$) }}}
{\vspace{0.5cm}

It is well known that the irreducible basis (IRB) of SU($mn$) $\supset$
SU($m$) $\times$ SU($n$) can be constructed from the IRBs of SU($m$)
and SU($n$) by using the CG coefficients of permutation group.
\begin{equation}
 \left| \begin{array}{c} [\nu] \\ 
 m, \beta [\nu_1]W_1[\nu_1]W_2 \end{array} \right\rangle
 = \sum_{m_1,m_2} 
 C^{\nu_{\beta} m}_{\nu_1 m_1, \nu_2 m_2}
 \left| \begin{array}{c} [\nu_1] \\ 
 m_1 W_1  \end{array} \right\rangle
 \left| \begin{array}{c} [\nu_2] \\ 
 m_2 W_2  \end{array} \right\rangle .
\end{equation}
where $\beta$ is the multiplicity index in the coupling $[\nu_1] \times
[\nu_2] \rightarrow [\nu]$, $m_1,m_2$ and $m$ are Yamanouchi labels.

Based on Eq.(4), we can write down multiquark wave functions (for
6-quark color singlets) in the different schemes. The steps to achieve
this are:

{\noindent 1. FS scheme: }

(1) $ [f] \times [\sigma_J] \rightarrow [\mu] $
\begin{equation}
 \left| \begin{array}{c} [\mu] \\ 
 m_{\mu}, \beta [f]W_f[\sigma_J]W_J \end{array} \right\rangle
 = \sum_{m_f,m_J} 
 C^{\mu_{\beta} m_{\mu} }_{f m_f, \sigma_J m_J}
 \left| \begin{array}{c} [f] \\ 
 m_f W_f  \end{array} \right\rangle
 \left| \begin{array}{c} [\sigma_J] \\ 
 m_J W_J  \end{array} \right\rangle ;
\end{equation}

(2) $ [\sigma] \times [\mu] \rightarrow [{\tilde \nu}] $
\begin{eqnarray}
 \left| \begin{array}{c} [{\tilde \nu}] \\ 
 m_{\tilde \nu}, \alpha [\mu]\beta [f]W_f[\sigma_J]W_J[\sigma]W_c 
 \end{array} \right\rangle
 & = & \sum_{m_f,m_J,m_c,m_{\mu} } 
 C^{{\tilde \nu_{\alpha}} m_{{\tilde \nu}}}_{\sigma m_c, \mu m_{\mu}}
 C^{\mu_{\beta} m_{\mu} }_{f m_f, \sigma_J m_J} \nonumber \\
 & & \left| \begin{array}{c} [\sigma] \\ 
 m_c W_c  \end{array} \right\rangle
 \left| \begin{array}{c} [f] \\ 
 m_f W_f  \end{array} \right\rangle
 \left| \begin{array}{c} [\sigma_J] \\ 
 m_J W_J  \end{array} \right\rangle ;
\end{eqnarray}

(3) $ [\nu] \times [{\tilde \nu}] \rightarrow [1^6] $
\begin{equation}
 \Phi^{FS} = \sum_{\mbox{all } m} 
 C^{[1^6]1}_{\nu m_{\nu}, {\tilde \nu} m_{{\tilde \nu}} }
 C^{{\tilde \nu_{\alpha}} m_{\tilde \nu}}_{\sigma m_c, \mu m_{\mu}}
 C^{\mu_{\beta} m_{\mu} }_{f m_f, \sigma_J m_J}
 \left| \begin{array}{c} [\nu] \\ 
 m_{\nu} W_{\nu}  \end{array} \right\rangle
 \left| \begin{array}{c} [\sigma] \\ 
 m_c W_c  \end{array} \right\rangle
 \left| \begin{array}{c} [f] \\ 
 m_f W_f  \end{array} \right\rangle
 \left| \begin{array}{c} [\sigma_J] \\ 
 m_J W_J  \end{array} \right\rangle .
\end{equation}

{\noindent 2. CS scheme: }

Following the same steps, the multiquark wave function can be written
as
\begin{equation}
 \Phi^{CS} = \sum_{\mbox{all } m} 
 C^{[1^6]1}_{\nu m_{\nu}, {\tilde \nu} m_{{\tilde \nu}} }
 C^{{\tilde \nu_{\alpha'}} m_{\tilde \nu}}_{\mu' m_{\mu'}, f m_f}
 C^{\mu'_{\beta'} m_{\mu'} }_{\sigma m_c, \sigma_J m_J}
 \left| \begin{array}{c} [\nu] \\ 
 m_{\nu} W_{\nu}  \end{array} \right\rangle
 \left| \begin{array}{c} [f] \\ 
 m_f W_f  \end{array} \right\rangle
 \left| \begin{array}{c} [\sigma] \\ 
 m_c W_c  \end{array} \right\rangle
 \left| \begin{array}{c} [\sigma_J] \\ 
 m_J W_J  \end{array} \right\rangle .
\end{equation}

The multiquark wave functions in different schemes are related by a
unitary transformation, so we have
\begin{equation}
 \Phi_i^{CS} = \sum_j C_{ij} \Phi_j^{FS}.
\end{equation}
where $i,j$ represent intermediate quantum numbers ${\tilde
\nu_{\alpha'}}, \mu'_{\beta'}, {\tilde \nu_{\alpha}}, \mu_{\beta}$.
Substituting Eqs.(7) and (8) into Eq.(9) and making use of the
orthonormality of bases and unitary conditions of CG coefficients, we
directly derive the expression for the transformation coefficients,
$C_{ij}$:
\begin{equation}
 C_{ij} = \sum_{m_c, m_J, m_{\mu}, m_f} 
 C^{{\tilde \nu_{\alpha'}} m_{\tilde \nu}}_{\mu' m_{\mu'}, f m_f}
 C^{\mu'_{\beta'} m_{\mu'} }_{\sigma m_c, \sigma_J m_J}
 C^{{\tilde \nu_{\alpha}} m_{\tilde \nu}}_{\sigma m_c, \mu m_{\mu}}
 C^{\mu_{\beta} m_{\mu} }_{f m_f, \sigma_J m_J}.
\end{equation}
Comparing Eq.(10) with the definition of the Racah coefficients of the
permutation group, we see that the $C_{ij}$ are just the Racah
coefficients except for a phase factor,
\begin{equation}
 C_{ij} = U(\sigma \sigma_J {\tilde \nu} f; \mu' \mu)
 ^{{\beta}'{\alpha}'}_{{\beta}{\alpha}} \epsilon (\sigma_J f \mu_{\beta})
\end{equation}
where $\alpha$, $\alpha'$, $\beta$, and $\beta'$ are multiplicity
indices appearing in the coupling, $\sigma \times \mu \rightarrow
{\tilde \nu}, \mu' \times f \rightarrow {\tilde \nu}, f \times \sigma_J
\rightarrow \mu, \sigma \times \sigma_J \rightarrow \mu'$,
respectively, and the phase factor $\epsilon (\sigma_J f \mu_{\beta})$
comes from the interchange of the order of $f$ and $\sigma_J$ in the
coupling $f \times \sigma_J \rightarrow \mu_{\beta}$.

In order to obtain the {\it rep} transformation coefficients, the Racah
coefficients of the permutation group are needed. (The phase factor
$\epsilon (\sigma_J f \mu_{\beta})$ is already available$^5$). We are
unaware of any publication of the complete Racah coefficients for the
permutation group (up to S(6)). Vanagas has given algebraic expressions
for some simple cases$^6$.  Because of the phase choices available for
the SU($n$) CG coefficients, one has to choose a systematic and phase
consistent method to calculate the Racah coefficients, especially for
physical applications.

We use the genealogical method to calculate the Racah coefficients of
S($n$) ($n\leq 6$) based on the existing phase consistent isoscalar
factors (ISF) of S($n)\supset$ S($n-1$) ($n\leq 6$). We start from the
trivial S(2) Racah coefficients, by using the formula$^7$
\begin{eqnarray}
 U(\nu_1 \nu_2 \nu \nu_3;\nu_{12} \nu_{23} )^{\beta_{12} \beta}
  _{\beta_{23} \beta'} & = & \sum_ {\nu'_1 \nu'_2 \nu'_3 \nu'_{12} \nu'_{23}
  \theta_{12} \theta_{23} \theta \theta'}^{\mbox{fixed } \nu'}
 U(\nu'_1 \nu'_2 \nu' \nu'_3;\nu'_{12} \nu'_{23} )^{\theta_{12} \theta}
  _{\theta_{23} \theta'} \nonumber \\
 & & 
 C^{\nu_{12} \beta_{12},\nu'_{12} \theta_{12} }_{\nu_1 \nu'_1,\nu_2 \nu'_2}
 C^{\nu \beta,\nu' \theta }_{\nu_{12} \nu'_{12},\nu_3 \nu'_3}
 C^{\nu_{23} \beta_{23},\nu'_{23} \theta_{23} }_{\nu_2 \nu'_2,\nu_3 \nu'_3}
 C^{\nu \beta',\nu' \theta' }_{\nu_{1} \nu'_{1},\nu_{23} \nu'_{23} }
\end{eqnarray}
where $\nu$'s and $\beta$'s are Yamanouchi labels and multiplicity
indices of S($n$), and $\nu'$'s and $\theta$'s are those of S($n-1$).
$C^{\nu_{12} \beta_{12},\nu'_{12} \theta_{12} }_{\nu_1 \nu'_1,\nu_2
\nu'_2}, \cdots$ are the S($n$) $\supset$ S($n-1$) isoscalar factors.
The Racah coefficients of S($n$) can be obtained sequentially in $n$.

\vspace{1cm}

{\noindent {\large {\bf III. Results }}}
\vspace{0.5cm}

We have carried out the calculation of the Racah coefficients of S($n$)
for $n\leq 6$, based on this genealogical method. The Racah
coefficients of S(6) needed in the transformation between the FS scheme
and the CS scheme are given in Table 1.  The {\it rep} transformation
coefficients between the FS and CS schemes in the $u$, $d$, and $s$
3-flavor world are obtained with the aid of parts of the table of phase
factors given by Chen {\em et al}$^5$, which are shown in Table 2.

We reproduce Tables VII and VIII of Stancu$^4$ for the transformation
between the FS scheme and the CS scheme with the exception of the
absolute phase of some bases. (For ($YIJ$)=(201), the phases of
[42][3111] and [42][2111] for the CS scheme should be negative. For
($YIJ$)=(200), no phases need be changed).  We also expand the physical
bases in terms of symmetry bases in the CS scheme. Tables 3--5 show the
results for ($YIJ$)=(210), (201) and (000), respectively.

Here, we want to emphasize that the importance of phase consistency.
Our result for ($YIJ$)=(201) is different from that of Table IX of
Stancu.  This is due to a phase inconsistency in the calculation of
Stancu, as we now show.  In expressing the physical bases in terms of
symmetry bases in the CS scheme through the expression of physical
bases in the FS scheme and the transformation between the FS scheme and
CS scheme, we need two successive transformations. In either step, the
absolute phase choice of each basis can be chosen arbitrarily, but one
has to use the same phase for each of the intermediate FS coupling
symmetry bases to calculate the {\it rep} transformation coefficients
in either step.

Since we can reproduce Tables VII and VIII of Stancu, the phase choice
for the FS coupling symmetry bases of Stancu is the same as ours.
However the phase choice for the FS coupling symmetry bases of Harvey
is different$^{3}$. Stancu directly combined Harvey's transformation
Table XI with her Table VII to obtain her Table IX.  The phase
inconsistency in making these two transformations produces the
incorrect entries in Table IX. This can be seen by the following
general argument.

For the nucleon, $N$, the CS symmetry must be [$\mu_i$]=[21] due to the
color singlet [1$^3$] and 1/2 spin [21] symmetry properties. Similarly,
for the $NN$ physical basis, the possible CS symmetries, according to
the Littlewood rule, are [42],[411],[33], [321],[2211] and [3111]. The
symmetry [21111] cannot appear in the coupling of [21]$\times$[21], and
so the entry in the last column of the first row in the Table IX of
Stancu should be zero. Similar arguments can be applied to the 2nd, 3rd
and 4th columns of the 2nd row in the same Table; all these entries
should be zero.

In conclusion, we emphasize again that one must be extremely careful in
quoting results from different papers because of the lack of unified
phase conventions.

\vspace{1cm}

{\noindent{ \large {\bf ACKNOWLEDGMENTS}} }
\vspace{0.5cm}

This work is supported by NSFC, the fundamental research project of the
SSTC, the graduate study fund of the SEC and DOE of US and the Natural 
Science fund of Jiangsu Province.

\vspace{0.3in}

\begin{description}
\item[{1.}]  M. Harvey, {\em Nucl. Phys.} {\bf A 352} (1981) 301. 

\item[{2.}] J.Q.Chen {\em et al}., {\em Nucl. Phys.} {\bf A 393} (1983) 122.

\item[{3.}]  J. L. Ping, F. Wang and T. Goldman, {\em Chin. J. Nucl. Phys.} 
 {\bf 16} (1994) 26. \\ \hspace{-0.1in} F. Wang, J. L. Ping and T. Goldman, 
 {\em Phys. Rev.} {\bf C51} (1995) 1648.

\item[{4.}]  Fl. Stancu, {\em Phys. Rev.} {\bf C39} (1989) 2030.

\item[{5.}]  J. Q. Chen {\em et al}., {\em Tables of the SU(mn)$\supset $ 
SU(m)$\times $SU(n) Coefficients of Fractional Parentage} (World Scientific, 
Singapore, 1991).

\item[{6.}]  Vanagas, {\em Group theory and nuclear spectroscopy }
(Minitis, USSR, 1967).

\item[{7.}]  J. Q. Chen, {\em The Group Representation Theory for
 Physicists} (World Scientific, Singapore, 1991) p254.

\end{description}

\newpage

\begin{small}
\begin{center}
Table. 1. 
\end{center}
The Racah coefficients $U(\nu_1 \nu_2 \nu \nu_3;\nu_{12} \nu_{23} 
  )^{\beta_{12} \beta}_{\beta_{23} \beta'}$ of S(6) with [$\nu_1$]=[222], 
  $\beta_{12}=1$ and $\beta'=1$. The meaning of the table heading is as
  follows:
\begin{center}
[$\nu$]~~~~~~~~~~~~~~~~~~~~~~

\begin{tabular}{|l|l|c} \hline 
[$\nu_2];[\nu_3$] & & $[\nu_{23}]_{\beta_{23}}$ \\ \cline{2-3}
 & [$\nu_{12}]_{\beta}$  &  \\ \hline
\end{tabular}
\end{center}
where the symmetry originally denoted by Young diagrams is denoted by 
numbers defined in table 6. All the entries represent the 
square of the Racah coefficients and a minus sign signifies a negative
coefficient value (This applies to all the Tables of this paper).
\end{small}

\vspace{0.2in}

[$\nu$] = [$1^6$] 

\begin{center}
\begin{tabular}{|l|l|c|} \hline 
1;5 & & 5 \\ \cline{2-3}
 & 7 &  1 \\ \hline
\end{tabular}
\hspace{5mm}
\begin{tabular}{|l|l|c|} \hline 
2;3 & & 5 \\ \cline{2-3}
 & 9 & 1 \\ \hline
\end{tabular}
\hspace{5mm}
\begin{tabular}{|l|l|c|} \hline 
2;6 & & 5 \\ \cline{2-3}  
 & 6 & 1  \\ \hline
\end{tabular}
\hspace{5mm}
\begin{tabular}{|l|l|c|} \hline 
3;2 & & 5 \\ \cline{2-3}  
 & 10 & 1 \\ \hline
\end{tabular}

\begin{tabular}{|l|l|c|} \hline 
3;4 & & 5 \\ \cline{2-3}
 & 8 & -1 \\ \hline
\end{tabular}
\hspace{5mm}
\begin{tabular}{|l|l|c|} \hline 
3;5 & & 5 \\ \cline{2-3}
 & 7 & -1 \\ \hline
\end{tabular}
\hspace{5mm}
\begin{tabular}{|l|l|c|} \hline 
3;6 & & 5 \\ \cline{2-3}
 & 6 & -1 \\ \hline
\end{tabular}
\hspace{5mm}
\begin{tabular}{|l|l|c|} \hline 
3;9 & & 5 \\ \cline{2-3}
 & 3 & -1 \\ \hline
\end{tabular}

\begin{tabular}{|l|l|c|} \hline 
5;1 & & 5 \\ \cline{2-3}
 & 11 & 1 \\ \hline
\end{tabular}
\hspace{5mm}
\begin{tabular}{|l|l|c|} \hline 
5;3 & & 5 \\ \cline{2-3}
 & 9 &  -1 \\ \hline
\end{tabular}
\hspace{5mm}
\begin{tabular}{|l|l|c|} \hline 
5;7 & & 5 \\ \cline{2-3}
 & 5 & -1 \\ \hline
\end{tabular}
\hspace{5mm}
\begin{tabular}{|l|l|c|} \hline 
5;8 & & 5 \\ \cline{2-3}
 & 4 & -1 \\ \hline
\end{tabular}

\end{center}

[$\nu$] = [2$1^5$]

\begin{center}
\begin{tabular}{|l|l|c|} \hline 
1;3 & & 3 \\ \cline{2-3}
 & 7 & 1 \\ \hline
\end{tabular}
\hspace{5mm}
\begin{tabular}{|l|l|c|} \hline 
1;6 & & 6 \\ \cline{2-3}
 & 7 & 1 \\ \hline
\end{tabular}
\hspace{5mm}
\begin{tabular}{|l|l|c|} \hline 
2;2 & & 3 \\ \cline{2-3}
 & 9 & 1 \\ \hline
\end{tabular}
\hspace{5mm}
\begin{tabular}{|l|l|cc|} \hline 
2;3 & & 3 & 6 \\ \cline{2-4}
 & 6 & 16/25 & -9/25 \\ 
 & 9 & ~9/25 & 16/25 \\ \hline
\end{tabular}
\hspace{5mm}
\begin{tabular}{|l|l|cc|} \hline 
2;4 & & 3 & 6 \\ \cline{2-4}
 & 6 & -24/25 &  ~-1/25 \\ 
 & 9 & ~~1/25 &  -24/25 \\ \hline
\end{tabular}
\hspace{5mm}
\begin{tabular}{|l|l|cc|} \hline 
2;5 & & 6 & 3 \\ \cline{2-4}
 & 6 & 1 & 0 \\
 & 9 & 0 & -1 \\ \hline
\end{tabular}
\hspace{5mm}
\begin{tabular}{|l|l|ccc|} \hline 
2;6 & & 3 & 6$_1$ & 6$_2$  \\ \cline{2-5}
 & 6$_1$ & -18/25 & ~-1/25 & ~~6/25 \\
 & 6$_2$ & ~-3/25 & ~-6/25 &  -16/25 \\
 & 9 & ~~4/25 &  -18/25 & ~~3/25 \\ \hline
\end{tabular}
\hspace{5mm}
\begin{tabular}{|l|l|c|} \hline 
2;7 & & 6 \\ \cline{2-3}
 & 6 & -1  \\ \hline
\end{tabular}
\hspace{5mm}
\begin{tabular}{|l|l|c|} \hline 
2;8 & & 6 \\ \cline{2-3}
 & 6 & 1 \\ \hline
\end{tabular}
\hspace{5mm}
\begin{tabular}{|l|l|c|} \hline 
2;9 & & 6 \\ \cline{2-3}
 & 6 & 1  \\ \hline
\end{tabular}
\hspace{5mm}
\begin{tabular}{|l|l|c|} \hline 
3;1 & & 3 \\ \cline{2-3}
 & 10 &  1  \\ \hline
\end{tabular}
\hspace{5mm}
\begin{tabular}{|l|l|cc|} \hline 
3;2 & & 3 & 6 \\ \cline{2-4}
 & 8 & -4/5 &  1/5   \\
 & 10 & ~1/5 &  4/5  \\ \hline
\end{tabular}
\hspace{5mm}
\begin{tabular}{|l|l|cccc|} \hline 
3;3 & & 3$_1$ & 3$_2$ & 6$_1$ & 6$_2$  \\ \cline{2-6}
 & 6 & -16/45 &  -16/45 &  1/90 &  5/18 \\
 & 7 &  ~1/9  &  -4/9   &  2/9  &  -2/9 \\
 & 8 & ~-4/45 &   -4/45 &  -49/90 &   -5/18 \\
 & 10 &  ~4/9 &   -1/9  &  -2/9  &   2/9 \\ \hline
\end{tabular}
\hspace{5mm}
\begin{tabular}{|l|l|ccc|} \hline 
3;4 & & 3 & 6$_1$ & 6$_2$ \\ \cline{2-5}
 & 6 & -8/15 &   -2/15 &   -1/3   \\
 & 8 &  4/15 &  1/15 &   -2/3 \\
 & 10 & -1/5 &  4/5 &   0 \\ \hline
\end{tabular}
\hspace{5mm}
\begin{tabular}{|l|l|c|} \hline 
3;5 & & 6 \\ \cline{2-3}
 & 6 & -1 \\ \hline
\end{tabular}
\hspace{5mm}
\begin{tabular}{|l|l|ccccc|} \hline 
3;6 & & 3$_1$ & 3$_2$ & 6$_1$ & 6$_2$  & 6$_3$ \\ \cline{2-7}
 & 3 &  9/40  &  9/40 &   0   &  1/20  &   1/2  \\
 & 6$_1$ & -1/5  &  1/5  &  4/25 & -2/5   &   1/25 \\
 & 6$_2$ & -3/10 & -1/30 &  -32/75 &   0 &    6/25 \\
 & 7 &  1/8  &   1/8  &   -2/5  & -1/4   &  -1/10 \\
 & 8 & -3/20 &   5/12 &   -1/75 &  3/10  &  -3/25 \\ \hline
\end{tabular}
\hspace{5mm}
\begin{tabular}{|l|l|cc|} \hline 
3;7 & & 3 & 6 \\ \cline{2-4}
 & 3 &  9/25 &  -16/25 \\
 & 6 & 16/25 &  9/25 \\ \hline
\end{tabular}
\hspace{5mm}
\begin{tabular}{|l|l|ccc|} \hline 
3;8 & & 3 & 6$_1$ & 6$_2$ \\ \cline{2-5}
 & 3 & -18/25 &  1/5  &   2/25 \\
 & 6 & -16/75 &   -2/15 &  -49/75 \\
 & 8 & -1/15  &  -2/3  &   4/15 \\ \hline
\end{tabular}
\hspace{5mm}
\begin{tabular}{|l|l|cc|} \hline 
3;9 & & 6$_1$ & 6$_2$  \\ \cline{2-4}
 & 3 &  1/10 & -9/10 \\
 & 6 &  9/10 &  1/10 \\ \hline
\end{tabular}
\hspace{5mm}
\begin{tabular}{|l|l|c|} \hline 
3;10 & & 6 \\ \cline{2-3}
 & 3 & -1  \\ \hline
\end{tabular}
\hspace{5mm}
\begin{tabular}{|l|l|cc|} \hline 
5;2 & & 3 & 6 \\ \cline{2-4}
 & 9 & -16/25 &  9/25  \\
 & 11 &  9/25 & 16/25  \\ \hline
\end{tabular}
\hspace{5mm}
\begin{tabular}{|l|l|c|} \hline 
5;3 & & 6 \\ \cline{2-3}
 & 9 &  1  \\ \hline
\end{tabular}
\hspace{5mm}
\begin{tabular}{|l|l|cc|} \hline 
5;4 & & 3 & 6 \\ \cline{2-4}
 & 4 &  3/5 &  2/5 \\
 & 9 & -2/5 &  3/5 \\ \hline
\end{tabular}
\hspace{5mm}
\begin{tabular}{|l|l|c|} \hline 
5;5 & & 3 \\ \cline{2-3}
 & 9 &  1 \\ \hline
\end{tabular}
\hspace{5mm}
\begin{tabular}{|l|l|ccc|} \hline 
5;6 & & 3 & 6$_1$ & 6$_2$ \\ \cline{2-5}
 & 4 &  3/10 &  1/10 &  3/5 \\
 & 5 & -9/20 & -3/20 &  2/5 \\
 & 9 & -1/4 &  3/4 & 0 \\ \hline
\end{tabular}
\hspace{5mm}
\begin{tabular}{|l|l|c|} \hline 
5;8 & & 6 \\ \cline{2-3}
 & 4 &  1 \\ \hline
\end{tabular}
\hspace{5mm}
\begin{tabular}{|l|l|cc|} \hline 
5;9 & & 3 & 6 \\ \cline{2-4}
 & 4 &  4/5 & -1/5 \\
 & 5 & -1/5 & -4/5 \\ \hline
\end{tabular}
\hspace{5mm}
\begin{tabular}{|l|l|c|} \hline 
5;10 & & 6 \\ \cline{2-3}
 & 4 & -1 \\ \hline
\end{tabular}
\end{center}

[$\nu$] = [2211]

\begin{center}
\begin{tabular}{|l|l|c|} \hline 
1;2 & & 2 \\ \cline{2-3}
 & 7 &  1 \\ \hline
\end{tabular}
\hspace{5mm}
\begin{tabular}{|l|l|c|} \hline 
1;4 & & 4 \\ \cline{2-3}
 & 7 &  1 \\ \hline
\end{tabular}
\hspace{5mm}
\begin{tabular}{|l|l|c|} \hline 
1;5 & & 5 \\ \cline{2-3}
 & 7 &  1 \\ \hline
\end{tabular}
\hspace{5mm}
\begin{tabular}{|l|l|c|} \hline 
1;6 & & 6 \\ \cline{2-3}
 & 7 &  1 \\ \hline
\end{tabular}
\hspace{5mm}
\begin{tabular}{|l|l|c|} \hline 
1;9 & & 9 \\ \cline{2-3}
 & 7 &  1 \\ \hline
\end{tabular}
\hspace{5mm}
\begin{tabular}{|l|l|c|} \hline 
2;1 & & 2 \\ \cline{2-3}
 & 9 &  1 \\ \hline
\end{tabular}
\hspace{5mm}
\begin{tabular}{|l|l|cc|} \hline 
2;2 & & 2 & 4 \\ \cline{2-4}
 & 6 & -4/5 &  1/5 \\
 & 9 & ~1/5 &  4/5  \\ \hline
\end{tabular}
\hspace{5mm}
\begin{tabular}{|l|l|cccc|} \hline 
2;3 & & 2 & 4 & 5 & 6 \\ \cline{2-6}
 & 6$_1$ & -2/9 &  -49/90 & -2/9 & -1/90 \\
 & 6$_2$ &  2/9 & -5/18 &  2/9 & -5/18 \\
 & 9$_1$ & -4/9 &  4/45 &  1/9 &  -16/45 \\
 & 9$_2$ & -1/9 & -4/45 &  4/9 & 16/45 \\ \hline
\end{tabular}
\hspace{5mm}
\begin{tabular}{|l|l|ccc|} \hline 
2;4 & & 2 & 4 & 6 \\ \cline{2-5}
 & 6$_1$ &  4/5 &  1/15 &  2/15  \\
 & 6$_2$ & 0 &  2/3 & -1/3    \\
 & 9 & -1/5 &  4/15 &  8/15 \\ \hline
\end{tabular}
\hspace{5mm}
\begin{tabular}{|l|l|c|} \hline 
2;5 & & 6 \\ \cline{2-3}
 & 6 &  1 \\ \hline
\end{tabular}
\hspace{5mm}
\begin{tabular}{|l|l|ccccc|} \hline 
2;6 & & 4 & 5 & 6$_1$ & 6$_2$ & 9 \\ \cline{2-7}
 & 6$_1$ &  1/75 & -2/5 & -4/25  & -32/75 & 0  \\
 & 6$_2$ &  3/10 &  1/4 & -2/5 & 0 & -1/20 \\
 & 6$_3$ & -3/25 &  1/10 &  1/25 & -6/25 & -1/2 \\
 & 9$_1$ &  3/20 &  1/8 &  1/5 & -3/10 &  9/40 \\
 & 9$_2$ & -5/12 &  1/8 & -1/5 & -1/30 &  9/40 \\ \hline
\end{tabular}
\hspace{5mm}
\begin{tabular}{|l|l|cc|} \hline 
2;7 & & 6 & 9 \\ \cline{2-4}
 & 6 &  9/25 & 16/25 \\
 & 9 & 16/25 & -9/25 \\ \hline
\end{tabular}
\hspace{5mm}
\begin{tabular}{|l|l|ccc|} \hline 
2;8 & & 4 & 6 & 9 \\ \cline{2-5}
 & 6$_1$ &  2/3 & -2/15 &  1/5 \\
 & 6$_2$ &  4/15 & 49/75 & -2/25 \\
 & 9 & -1/15 & 16/75 & 18/25 \\ \hline
\end{tabular}
\hspace{5mm}
\begin{tabular}{|l|l|cc|} \hline 
2;9 & & 6 & 9 \\ \cline{2-4}
 & 6$_1$ &  9/10 & -1/10 \\
 & 6$_2$ & -1/10 & -9/10 \\ \hline
\end{tabular}
\hspace{5mm}
\begin{tabular}{|l|l|c|} \hline 
2;10 & & 9 \\ \cline{2-3}
 & 6 & -1  \\ \hline
\end{tabular}
\hspace{5mm}
\begin{tabular}{|l|l|cccc|} \hline 
3;2 & & 2 & 4 & 5 & 6 \\ \cline{2-6}
 & 6 & 20/81 & -25/81 & 20/81  & -16/81 \\
 & 7 & -16/81 & 20/81 & 25/81  & -20/81 \\
 & 8 & 20/81 & 16/81 & 20/81 & 25/81 \\
 & 10 & 25/81 & 20/81  & -16/81  & -20/81 \\ \hline
\end{tabular}
\hspace{5mm}
\begin{tabular}{|l|l|cccc|} \hline 
3;3 & & 2 & 4 & 6$_1$ & 6$_2$ \\ \cline{2-6}
 & 6$_1$ &  2/9 &  1/18 & 25/36 &  1/36  \\
 & 6$_2$ & -2/9 &  1/2 &  1/36 & -1/4 \\
 & 8 &  4/9 & 0 & -1/18 & -1/2   \\
 & 10 &  1/9 &  4/9 & -2/9 &  2/9 \\ \hline
\end{tabular}
\hspace{5mm}
\begin{tabular}{|l|l|ccccccc|} \hline 
3;4 & & 2 & 4$_1$ & 4$_2$ & 5 & 6$_1$ & 6$_2$ & 9 \\ \cline{2-9}
 & 3 &  2/5 &  1/16 &  5/16 &  1/16 & 0 & 4/25 &  1/400 \\
 & 6$_1$ &  4/81 &  5/18 &  1/162 & -40/81 & 4/81 & -10/81 & 0 \\
 & 6$_2$ & 128/405 & -1/9 & -5/81 &  4/81 & 10/81 & -361/2025 & -4/25 \\
 & 7 & 10/81 &  1/16 & -245/1296 & 25/1296 & -40/81 & -4/81 &  1/16 \\
 & 8$_1$ & 0 & 25/144 & -5/16 &  1/16 & 5/18 & 1/9 &  1/16 \\
 & 8$_2$ &  8/81 & -5/16 & -25/1296 & -245/1296 & 1/162 & 5/81 &  5/16 \\
 & 10 &  1/81 & 0 & -8/81 & -10/81 & -4/81 & 128/405 & -2/5 \\ \hline
\end{tabular}
\hspace{5mm}
\begin{tabular}{|l|l|ccccc|} \hline 
3;5 & & 2 & 4 & 5 & 6 & 9 \\ \cline{2-7}
 & 3 &  1/5 &  1/40 &  5/16 & -4/25 & -121/400 \\
 & 6 &-64/405 & -8/405 &  -20/81 & -841/2025 & -4/25 \\
 & 7 & 25/81 & 25/648 & -121/1296 &  -20/81 &  5/16 \\
 & 8 & -2/81 &  289/324 &  -25/648 &  8/405 & -1/40 \\
 & 10 &-25/81 &  2/81 & 25/81 &  -64/405 &  1/5  \\ \hline
\end{tabular}
\hspace{5mm}

\begin{tiny}
\begin{tabular}{|l|l|ccccccccc|} \hline 
3;6 & & 2 & 4$_1$ & 4$_2$ & 5 & 6$_1$ & 6$_2$ & 6$_3$ & 9$_1$ & 9$_2$ 
   \\ \cline{2-11}
 & 3$_1$ &  1/8 &  1/8 &  1/5 & -1/8 &  9/125 & -49/200 & -49/500 & -1/100 
   & 0 \\
 & 3$_2$ &  1/8 & -1/8 & 0 & -1/8 &  1/5 &  1/8 &  1/20 & 0 &  1/4 \\
 & 6$_1$ &  8/45 & 0 & 49/225 &  2/45 & -16/5625 & 32/1125 &  1444/5625 &  
   9/125 & -1/5 \\
 & 6$_2$ & -16/81 & -1/81 & 289/810 & -1/324 & -32/1125 & 16/2025 & 
   -242/10125 & 49/200 &  1/8 \\
 & 6$_3$ &  8/405 & 10/81 & -169/2025 & -169/810 & -1444/5625 & -242/10125 
   & 6889/50625 & 49/500 &  1/20 \\
 & 7 & 25/324 & -25/81 & -5/162 & -25/324 & -2/45 & -1/324 & -169/810 &  
    1/8 & -1/8 \\
 & 8$_1$ &-16/81 & -25/324 & 5/162 & -25/81 & 0 & -1/81 & 10/81 & -1/8 & 
   -1/8 \\
 & 8$_2$ &  5/162 & 5/162 &  4/81 & -5/162 & -49/225 & 289/810 & -169/2025 
    & -1/5 & 0 \\
 & 10 & -4/81 & 16/81 & -5/162 & -25/324 & 8/45 & 16/81 & -8/405 &  1/8 & 
   -1/8 \\ \hline
\end{tabular}
\end{tiny}

\begin{tabular}{|l|l|c|} \hline 
3;7 & & 6 \\ \cline{2-3}
 & 6 & -1  \\ \hline
\end{tabular}
\hspace{5mm}
\begin{tabular}{|l|l|cccc|} \hline 
3;8 & & 4 & 6$_1$ & 6$_2$ & 9 \\ \cline{2-6}
 & 3 & -2/5 & -1/25 & -2/5 & -4/25 \\
 & 6$_1$ & -2/45 & 196/225 & -2/45 &  1/25 \\
 & 6$_2$ &  4/9 &  2/45 & -1/9 & -2/5 \\
 & 8 & -1/9 &  2/45 &  4/9 & -2/5 \\ \hline
\end{tabular}
\hspace{5mm}
\begin{tabular}{|l|l|cccccc|} \hline 
3;9 & & 4 & 5 & 6$_1$ & 6$_2$ & 9$_1$ & 9$_2$ \\ \cline{2-8}
 & 3$_1$ &  9/40 &  1/16 &  -8/25 &   0 & -121/400  &  -9/100 \\
 & 3$_2$ & -1/10 &  1/4 &  1/50 &  1/2 & -9/100  &   1/25 \\
 & 6$_1$ & 16/45 &  2/9 &   49/900 &  1/36 &  8/25 &   -1/50 \\
 & 6$_2$ & 0  &   2/9  &  -1/36 &   -1/4 & 0 &  1/2 \\
 & 7 &  5/72 &  -25/144 &   -2/9 &  2/9  &   1/16  &   1/4 \\
 & 8 &  1/4  &  -5/72  &  16/45 &   0 & -9/40  &   1/10 \\ \hline
\end{tabular}
\hspace{5mm}
\begin{tabular}{|l|l|cc|} \hline 
3;10 & & 6 & 9 \\ \cline{2-4}
 & 3 & ~~9/25  &  16/25 \\
 & 6 & -16/25 &   ~9/25 \\ \hline
\end{tabular}
\hspace{5mm}
\begin{tabular}{|l|l|c|} \hline 
3;11 & & 9 \\ \cline{2-3}
 & 3 & -1 \\ \hline
\end{tabular}
\hspace{5mm}
\begin{tabular}{|l|l|c|} \hline 
5;1 & & 5 \\ \cline{2-3}
 & 9 &  1  \\ \hline
\end{tabular}
\hspace{5mm}
\begin{tabular}{|l|l|c|} \hline 
5;2 & & 6 \\ \cline{2-3}
 & 9 &  1  \\ \hline
\end{tabular}
\hspace{5mm}
\begin{tabular}{|l|l|ccccc|} \hline 
5;3 & & 2 & 4 & 5 & 6 & 9 \\ \cline{2-7}
 & 4 &  2/9 &   1/36 &   25/72 &   -8/45  &  -9/40 \\
 & 5 & -1/9 &   25/72 &  -25/144 &  -16/45 &   -1/80 \\
 & 9$_1$ &  1/9 &  -25/72 &  -49/144 &   -4/45 &   -9/80 \\
 & 9$_2$ &  4/9 &  1/18 &   -1/36 &   -1/45 &  9/20 \\
 & 11 &  1/9 &  2/9 &   -1/9 &   16/45  &  -1/5 \\ \hline
\end{tabular}
\hspace{5mm}
\begin{tabular}{|l|l|cc|} \hline 
5;4 & & 4 & 6 \\ \cline{2-4}
 & 4 & ~1/3  &  -2/3 \\
 & 9 & -2/3  &  -1/3 \\ \hline
\end{tabular}
\hspace{5mm}
\begin{tabular}{|l|l|ccccc|} \hline 
5;6 & & 2 & 4 & 6$_1$ & 6$_2$ & 9 \\ \cline{2-7}
 & 4$_1$ & 0  &  -5/12 &  1/3 &   0 &   -1/4 \\
 & 4$_2$ &  1/2 & 1/6  &   1/30 &   -1/5 &   -1/10 \\
 & 5 &  1/4  &  0   &  3/20  &   2/5  &   1/5 \\
 & 9$_1$ &  1/8 & -3/8 &   -3/40 &   -1/5 &  9/40 \\
 & 9$_2$ & -1/8 &  1/24 &   49/120 &   -1/5 &  9/40 \\ \hline
\end{tabular}
\hspace{5mm}
\begin{tabular}{|l|l|ccc|} \hline 
5;7 & & 4 & 5 & 9 \\ \cline{2-5}
 & 4 &  1/4 &   -5/8  &  -1/8  \\
 & 5 & -5/8 &   -1/16 &   -5/16 \\
 & 9 & -1/8 &   -5/16 &  9/16 \\ \hline
\end{tabular}
\hspace{5mm}
\begin{tabular}{|l|l|cccc|} \hline 
5;8 & & 4 & 5 & 6 & 9 \\ \cline{2-6}
 & 4$_1$ & -5/24 &  1/16  &   1/6 &   -9/16   \\
 & 4$_2$ &  3/8 &  5/16 &  3/10 &  1/80  \\
 & 5 &  3/8  &  -5/16 &   0 &   -5/16  \\
 & 9 & -1/24 &   -5/16 &  8/15 &  9/80 \\ \hline
\end{tabular}
\hspace{5mm}
\begin{tabular}{|l|l|c|} \hline 
5;9 & & 6 \\ \cline{2-3}
 & 4 & -1  \\ \hline
\end{tabular}
\hspace{5mm}
\begin{tabular}{|l|l|cc|} \hline 
5;10 & & 6 &  9 \\ \cline{2-4}
 & 4 & -1/5 &  -4/5 \\
 & 5 & -4/5 &  ~1/5 \\ \hline
\end{tabular}
\end{center}

[$\nu$] = [222]

\begin{center}
\begin{tabular}{|l|l|c|} \hline 
1;1 & & 1 \\ \cline{2-3}
 & 7 &  1 \\ \hline 
\end{tabular}
\hspace{5mm}
\begin{tabular}{|l|l|c|} \hline 
1;3 & & 3 \\ \cline{2-3}
 & 7 &  1 \\ \hline 
\end{tabular}
\hspace{5mm}
\begin{tabular}{|l|l|c|} \hline 
1;7 & & 7 \\ \cline{2-3}
 & 7 &  1 \\ \hline 
\end{tabular}
\hspace{5mm}
\begin{tabular}{|l|l|c|} \hline 
1;8 & & 8 \\ \cline{2-3}
 & 7 &  1 \\ \hline 
\end{tabular}
\hspace{5mm}
\begin{tabular}{|l|l|cc|} \hline 
2;2 & & 1 &  3 \\ \cline{2-4}
 & 6 & 16/25  &   9/25 \\
 & 9 &  9/25  & -16/25 \\ \hline 
\end{tabular}
\hspace{5mm}
\begin{tabular}{|l|l|c|} \hline 
2;3 & & 3 \\ \cline{2-3}
 & 6 &  1 \\ \hline 
\end{tabular}
\hspace{5mm}
\begin{tabular}{|l|l|cc|} \hline 
2;4 & & 3 & 8 \\ \cline{2-4}
 & 6 & -3/5   &  2/5 \\
 & 9 & ~2/5   &  3/5 \\ \hline 
\end{tabular}
\hspace{5mm}
\begin{tabular}{|l|l|c|} \hline 
2;5 & & 3 \\ \cline{2-3}
 & 9 &  1 \\ \hline 
\end{tabular}
\hspace{5mm}
\begin{tabular}{|l|l|ccc|} \hline 
2;6 & & 3 & 7 & 8 \\ \cline{2-5}
 & 6$_1$ &  3/4  &  -3/20 &   -1/10 \\
 & 6$_2$ & 0  &   2/5  &  -3/5 \\
 & 9 & -1/4 &   -9/20 &   -3/10 \\ \hline 
\end{tabular}
\hspace{5mm}
\begin{tabular}{|l|l|c|} \hline 
2;8 & & 8 \\ \cline{2-3}
 & 6 & -1 \\ \hline 
\end{tabular}
\hspace{5mm}
\begin{tabular}{|l|l|cc|} \hline 
2;9 & & 7 & 8 \\ \cline{2-4}
 & 6 &  4/5 &   ~1/5 \\
 & 9 &  1/5 &   -4/5 \\ \hline 
\end{tabular}
\hspace{5mm}
\begin{tabular}{|l|l|c|} \hline 
2;10 & & 8 \\ \cline{2-3}
 & 6 &  1 \\ \hline 
\end{tabular}
\hspace{5mm}
\begin{tabular}{|l|l|c|} \hline 
3;1 & & 3 \\ \cline{2-3}
 & 7 &  1 \\ \hline 
\end{tabular}
\hspace{5mm}
\begin{tabular}{|l|l|c|} \hline 
3;2 & & 3 \\ \cline{2-3}
 & 6 &  1 \\ \hline 
\end{tabular}
\hspace{5mm}
\begin{tabular}{|l|l|ccccc|} \hline 
3;3 & & 1 &  3$_1$ & 3$_2$ & 7 & 8 \\ \cline{2-7}
 & 3 &  1/5 &  9/80  &   9/20 &  1/80  &  -9/40 \\
 & 6 & 16/45 &   -4/45 &  1/45 &  -16/45  &   8/45 \\
 & 7 &  1/9  &  49/144 &   -1/36 &   25/144 &   25/72 \\
 & 8 &  2/9  & -25/72  &  -1/18 &  25/72 &   -1/36 \\
 & 10 &  1/9  &   1/9 &   -4/9  &  -1/9  &  -2/9 \\ \hline 
\end{tabular}
\hspace{5mm}
\begin{tabular}{|l|l|cc|} \hline 
3;4 & & 3 & 8 \\ \cline{2-4}
 & 6 & -1/3 &   -2/3  \\
 & 8 & -2/3 &  1/3  \\ \hline 
\end{tabular}
\hspace{5mm}
\begin{tabular}{|l|l|ccccc|} \hline 
3;6 & & 3$_1$ & 3$_2$ &  7 & 8$_1$ & 8$_2$ \\ \cline{2-7}
 & 3 & -9/40 &   -9/40 &   -1/5 &   -1/4 &   -1/10 \\
 & 6$_1$ &  3/40 &  -49/120  &  -3/20 &  1/3  &   1/30  \\
 & 6$_2$ &  1/5 &  1/5 &   -2/5 &   0 &   -1/5 \\
 & 8 &  3/8  &  -1/24 &   0  &  -5/12 &  1/6  \\
 & 10 &  1/8 &   -1/8 &  1/4 &   0 &   -1/2  \\ \hline 
\end{tabular}
\hspace{5mm}
\begin{tabular}{|l|l|ccc|} \hline 
3;7 & & 3 & 7 & 8 \\ \cline{2-5}
 & 3 &  9/16 &  5/16 &   -1/8 \\
 & 7 &  5/16 &   -1/16 &  5/8 \\
 & 8 & -1/8  &   5/8   &  1/4  \\ \hline 
\end{tabular}
\hspace{5mm}
\begin{tabular}{|l|l|cccc|} \hline 
3;8 & & 3 & 7 & 8$_1$ & 8$_2$ \\ \cline{2-6}
 & 3 &  9/80 &  5/16 &   -9/16 &   -1/80 \\
 & 6 &  8/15 &   0   &  1/6 &   -3/10  \\
 & 7 & -5/16 &  5/16  &   1/16  &  -5/16 \\
 & 8 &  1/24 &  3/8  &   5/24   &  3/8  \\ \hline 
\end{tabular}
\hspace{5mm}
\begin{tabular}{|l|l|c|} \hline 
3;9 & & 8 \\ \cline{2-3}
 & 6 &  1 \\ \hline 
\end{tabular}
\hspace{5mm}
\begin{tabular}{|l|l|cc|} \hline 
3;10 & & 7 & 8 \\ \cline{2-4}
 & 3 & -1/5 &   ~4/5 \\
 & 6 & -4/5 &   -1/5 \\ \hline 
\end{tabular}
\hspace{5mm}
\begin{tabular}{|l|l|c|} \hline 
5;2 & & 3 \\ \cline{2-3}
 & 9 &  1 \\ \hline 
\end{tabular}
\hspace{5mm}
\begin{tabular}{|l|l|ccc|} \hline 
5;4 & & 3 & 7 & 8 \\ \cline{2-5}
 & 4 & -3/8 &   -3/8 &   -1/4 \\
 & 5 &  9/16 &   -1/16 &   -3/8 \\
 & 9 &  1/16 &   -9/16 &  3/8 \\ \hline 
\end{tabular}
\hspace{5mm}
\begin{tabular}{|l|l|cccc|} \hline 
5;5 & & 1 & 3 & 7 & 8 \\ \cline{2-6}
 & 4 &  ~2/5  &   9/40  &  -1/8 &   -1/4  \\
 & 5 &  ~1/5  &   9/80  &   9/16 &  1/8  \\
 & 9 &  9/25  & -121/400 &   -9/80 &  9/40 \\
 & 11 &  ~1/25 &   -9/25 &  1/5 &   -2/5 \\ \hline 
\end{tabular}
\hspace{5mm}
\begin{tabular}{|l|l|cc|} \hline 
5;6 & & 3 & 8 \\ \cline{2-4}
 & 4 & -3/4 &  ~1/4 \\
 & 9 & -1/4 &  -3/4 \\ \hline 
\end{tabular}
\hspace{5mm}
\begin{tabular}{|l|l|c|} \hline 
5;8 & & 8 \\ \cline{2-3}
 & 4 & -1 \\ \hline 
\end{tabular}
\hspace{5mm}
\begin{tabular}{|l|l|ccc|} \hline 
5;9 & & 3 & 7 & 8 \\ \cline{2-5}
 & 4 & ~1/8  &   ~5/8  &  -1/4 \\
 & 5 & -5/16 &   -1/16 &   -5/8 \\
 & 9 & ~9/16 &   -5/16 &   -1/8 \\ \hline 
\end{tabular}
\hspace{5mm}
\begin{tabular}{|l|l|c|} \hline 
5;11 & & 7 \\ \cline{2-3}
 & 5 &  1 \\ \hline 
\end{tabular}

\end{center}

\begin{center}
Table 2. The triplets ($\sigma_J f \mu_{\beta}$) which have
 $\epsilon(\sigma_J f \mu_{\beta})=-1$.
\begin{tabular}{|c|l|c|l|c|l|c|l|c|l|} \hline
 [$\sigma_J];[f$] & [$\mu]_{\beta}$ & [$\sigma_J];[f$] & [$\mu]_{\beta}$ & 
 [$\sigma_J];[f$] & [$\mu]_{\beta}$ & [$\sigma_J];[f$] & [$\mu]_{\beta}$ &
 [$\sigma_J];[f$] & [$\mu]_{\beta}$  \\ \hline
 2;2 & 4  & 3;3 & 4     & 3;6 & 7     & 3;9 & 9$_2$ & 5;6 & 10  \\
 2;3 & 6  & 3;3 & 6$_2$ & 3;6 & 9$_1$ & 3;9 & 10    & 5;7 & 11 \\
 2;6 & 8  & 3;3 & 8     & 3;6 & 9$_2$ & 3;9 & 11    & &  \\
 2;6 & 7  & 3;4 & 8     & 3;6 & 10    & 5;5 & 8     & &  \\
 2;8 & 10 & 3;5 & 9     & 3;7 & 10    & 5;6 & 8     & &  \\ \hline
\end{tabular}
\end{center}
\begin{small}
$\epsilon(\sigma_J f \mu_{\beta})=\epsilon(f \sigma_J \mu_{\beta})$.
The cases $\epsilon(\sigma_J f \mu_{\beta})=1$ which occur when
$\sigma_J$ or $f$ are either totally symmetric or totally antisymmetric
are not shown. \\
All the triplets needed in this calculation, which are not listed in
Table 2, have $\epsilon(\sigma_J f \mu_{\beta})=1$.
\end{small}

\begin{center}
Table 3. The transformation between physical and symmetry 
~~~~~~bases in the CS scheme for (YIJ)=(210). The labels 
of the columns are the values of $ [\nu]$ $[\mu]_{\beta}$ $[f]$. ~~~~~~~~
\begin{tabular}{|c|cccccc|} \hline
 $B_1B_2$ & 1 9 3 & 3 5 3 & 3 4 3 & 3 9$_1$ 3 & 3 9$_2$ 3 & 3 11 3 \\ \hline
 $NN $ & $-\sqrt{1/9}$ & $\sqrt{1/2}$ & 
 $-\sqrt{1/4}$ & $ -\sqrt{1/36} $ & $\sqrt{1/9} $ & 
 0 \\
 $ \Delta\Delta $ & $\sqrt{4/45} $  & 0 & 0 &
 $ \sqrt{16/45} $ & $\sqrt{16/45}$ & $\sqrt{1/5}$  
 \\ \hline
\end{tabular}

\vspace{1cm}

Table 4. Same as Table 3, but for ($YIJ$)=(201).

\begin{tabular}{|c|cccccc|} \hline
$B_1B_2$ & 1 8 5 & 3 3 5 & 3 6 5 & 3 8 5 & 3 7 5 & 3 10 5 \\ \hline
 $NN $ & $ -\sqrt{1/9} $ & $\sqrt{9/20} $ & 
 $-\sqrt{16/45} $ & $ \sqrt{1/36} $ & $-\sqrt{1/18}$ 
 & 0 \\
 $ \Delta\Delta $ & $\sqrt{4/45} $ & 0 & 0 & 
 $ \sqrt{16/45} $ & 0 & $-\sqrt{5/9}$ \\ \hline
\end{tabular}

\vspace{1cm}

Table 5. Same as Table 3, but for ($YIJ$)=(000).
\begin{tabular}{|c|ccccccc|} \hline
$B_1B_2$ & 1 9 3 & 1 4 8 & 2 5 6 & 2 4 6 & 2 9 6 & 3 5 3 & 3 4 3 \\ \hline 
 $\overline{N\Xi} $ & 0 & 0 & $\sqrt{1/6}$ & $-\sqrt{1/4}$ 
 & $-\sqrt{5/36}$ & 0 & 0 \\
 $ \Sigma\Sigma $ & $\sqrt{1/360}$ & $-\sqrt{3/40}$ & 0 & 0 & 0 &
 $-\sqrt{1/80}$  & $\sqrt{1/160}$ \\
 $ \widetilde{N\Xi} $ & $\sqrt{1/30}$ & $\sqrt{1/10}$ & 0 & 0 & 
 0 & $-\sqrt{3/20}$ & $\sqrt{3/40}$  \\
 $ \Lambda\Lambda $ & $-\sqrt{3/40}$ & $\sqrt{1/40}$ & 0 & 0 & 
 0 & $\sqrt{27/80}$ & $-\sqrt{27/160}$  \\
 $ \Sigma^*\Sigma^* $ & $\sqrt{4/45}$ & 0 & 0 & 0 & 0 & 0 & 0 \\
 \hline
 & 3 9$_1$ 3 & 3 9$_2$ 3 & 3 11 3 & 3 5$_2$ 6 & 3 4 6 & 3 9$_1$ 6 & 3 9$_2$ 6
  \\ \hline
 $\overline{N\Xi} $ & 0 & 0 & 0 & 0  & 0 & 0 & 0 \\
 $ \Sigma\Sigma $ & $\sqrt{1/1440}$ & $-\sqrt{1/360}$ & 0 & 
 $\sqrt{1/10}$ & $\sqrt{3/20}$ & $\sqrt{3/40}$ & 
 $-\sqrt{3/40}$ \\
 $ \widetilde{N\Xi} $ & $\sqrt{1/120}$ & $-\sqrt{1/30}$ & 0 & 
 $\sqrt{1/10}$ & $\sqrt{1/20}$ & $\sqrt{1/40}$ & 
 $-\sqrt{1/40}$ \\
 $ \Lambda\Lambda $ & $-\sqrt{3/160}$ & $\sqrt{3/40}$ & 0 & 
 $\sqrt{1/10}$ & $\sqrt{1/20}$ & $\sqrt{1/40}$ &  
 $-\sqrt{1/40}$ \\
 $ \Sigma^*\Sigma^* $ & $\sqrt{16/45}$ & $\sqrt{16/45}$ & 
 $\sqrt{1/5}$ & 0 & 0 & 0 & 0 \\ \hline
 & 3 5 8 & 3 4 8 & 3 9 8 & 4 5 6 & 4 9 6 & & \\ \hline
 $\overline{N\Xi} $ & 0 & 0 & 0 & $-\sqrt{1/3}$ & 
 $-\sqrt{1/9}$ & &  \\
 $ \Sigma\Sigma $ & $-\sqrt{3/16}$  & $-\sqrt{3/160}$ &
 $-\sqrt{3/32}$ & 0 & 0 & & \\
 $ \widetilde{N\Xi} $ & $\sqrt{1/4}$  & $\sqrt{1/40}$ &
 $\sqrt{1/8}$ & 0 & 0 & & \\
 $ \Lambda\Lambda $ & $\sqrt{1/16}$  & $\sqrt{1/160}$ &
 $\sqrt{1/32}$ & 0 & 0 & & \\
 $ \Sigma^*\Sigma^* $  & 0 & 0 & 0 & 0 & 0 &  &  \\ \hline
\end{tabular}          
\end{center}

\begin{center}
Table 6. Column label coding, used in Table 1-5, for the Young diagrams.
 
\begin{tabular}{|ccccccccccc|} \hline
 1 & 2 & 3  & 4 & 5 & 6 & 7 & 8 & 9 & 10 & 11 \\ \hline
 [6] & [51] & [42] & [411] & [33] & [321] & [222] &  [3111] &  
 [2211] & [211111] & [111111] \\ \hline
\end{tabular}
\end{center}

\end{document}